\documentclass[10pt,british,english,aps,superscriptaddress,preprintnumbers,amsmath,nofootinbib,amssymb,altaffilletter]{revtex4}
\usepackage[T1]{fontenc}
\usepackage[latin9]{inputenc}
\usepackage{color}
\usepackage{babel}
\usepackage{float}
\usepackage{amsmath}
\usepackage{amssymb}
\usepackage{graphicx}
\usepackage{wasysym}
\usepackage{esint}
\usepackage[unicode=true,pdfusetitle,
bookmarks=true,bookmarksnumbered=false,bookmarksopen=false,
breaklinks=false,pdfborder={0 0 1},backref=false,colorlinks=false]
{hyperref}
\makeatletter

\@ifundefined{textcolor}{}
{%
	\definecolor{BLACK}{gray}{0}
	\definecolor{WHITE}{gray}{1}
	\definecolor{RED}{rgb}{1,0,0}
	\definecolor{GREEN}{rgb}{0,1,0}
	\definecolor{BLUE}{rgb}{0,0,1}
	\definecolor{CYAN}{cmyk}{1,0,0,0}
	\definecolor{MAGENTA}{cmyk}{0,1,0,0}
	\definecolor{YELLOW}{cmyk}{0,0,1,0}
}

\usepackage{babel}
\usepackage{babel}
\usepackage{babel}
\usepackage{babel}

\usepackage{babel}
\usepackage{breakurl}

\newcommand{\LF}{\left(}
\newcommand{\RF}{\right)}
\newcommand{\LT}{\left[}
\newcommand{\RT}{\right]}

\@ifundefined{textcolor}{}{%
	\definecolor{BLACK}{gray}{0}
	\definecolor{WHITE}{gray}{1}
	\definecolor{RED}{rgb}{1,0,0}
	\definecolor{GREEN}{rgb}{0,1,0}
	\definecolor{BLUE}{rgb}{0,0,1}
	\definecolor{CYAN}{cmyk}{1,0,0,0}
	\definecolor{MAGENTA}{cmyk}{0,1,0,0}
	\definecolor{YELLOW}{cmyk}{0,0,1,0}
}

\def\Fc{\mathcal{F}}

\def\Mc{\mathcal{M}}

\def\Oc{\mathcal{O}}
\def\Pc{\mathcal{P}}

\def\Rc{\mathcal{R}}

\usepackage{babel}
\usepackage{epsfig}\usepackage{mathrsfs}\usepackage{euscript}

\renewcommand{\[}{\begin{equation}}
\renewcommand{\]}{\end{equation}}

\makeatother

\begin{document}

\title{Analytic infinite derivative gravity, $R^2$-like inflation, quantum gravity and CMB}

\author{Alexey S. Koshelev}
\email{alex.koshelev@ubi.pt}

\selectlanguage{british}%

\address{Departamento de F\'{i}sica, Centro de Matem\'atica e Aplica\c{c}\~oes
	(CMA-UBI), Universidade da Beira Interior, 6200 Covilh\~a, Portugal}

\author{K. Sravan Kumar}
\email{sravan.korumilli@rug.nl,sravan.mph@icloud.com}

\selectlanguage{british}%

\address{Van  Swinderen  Institute,  University  of  Groningen,  9747  AG,  Groningen,  The  Netherlands}
\selectlanguage{english}%

\author{Alexei A. Starobinsky}
\email{alstar@landau.ac.ru}

\selectlanguage{british}%

\address{
	L. D. Landau Institute for Theoretical Physics RAS, Moscow 119334,
	Russian Federation}
\selectlanguage{english}%

\begin{abstract}
Emergence of $R^2$ inflation which is the best fit framework for CMB observations till date comes from the attempts to attack the problem of quantization of gravity which in turn have resulted in the trace anomaly discovery. Further developments in trace anomaly and different frameworks aiming to construct quantum gravity indicate an inevitability of non-locality in fundamental physics at small time and length scales. A natural question would be to employ the $R^2$ inflation as a probe for signatures of non-locality in the early Universe physics. Recent advances of embedding $R^2$ inflation in a string theory inspired non-local gravity modification provides very promising theoretical predictions connecting the non-local physics in the early Universe and the forthcoming CMB observations.
\\
\\
\\
\textit{This essay received an Honorable Mention in the 2020 Essay Competition of the Gravity Research
Foundation.}
\end{abstract}

\maketitle

\section*{Introducing the gravity modification}
According to recent studies \cite{Krasnikov:1987yj,Kuzmin:1989sp,Tomboulis:1997gg,Biswas:2005qr,Modesto:2011kw,Biswas:2011ar,Biswas:2016egy,Koshelev:2016xqb,Koshelev:2017tvv,Koshelev:2020foq} 
the following higher derivative quadratic in curvatures gravity action captures the long standing quests to build a much needed consistent theoretical and phenomenological path towards quantum gravity:
\begin{equation}\label{NC-action}
S=\int d^{4}x\sqrt{-g}\left(\frac{M_{p}^{2}}{2}R+
\frac{1}{2}
\bigg[R\mathcal{F}_{R}\left(\square_{s}\right)R+W_{\mu\nu\rho\sigma}\mathcal{F}_{W}\left(\square_{s}\right)W^{\mu\nu\rho\sigma}\bigg]\right)\,.
\end{equation}
Here the first term with $M_{p}$ being the reduced Planck mass is the canonical Einstein-Hilbert General Realtivity (GR) Lagrangian, $R$ as usual is the Ricci scalar, while the part in brackets is the higher derivative modification of GR. The crucial ingredient here is higher derivative formfactors $\Fc$ which further turn out to be analytic non-polynomial, i.e. essentially non-local, functions of the covariant d'Alembertian $\square$. Hereafter we use 
$\square_{s}={\square}/{\Mc_{s}^{2}}$ with $\Mc_{s}$ being the scale of non-locality and 
$W_{\mu\nu\rho\sigma}$ denotes the Weyl tensor.

The above gravity action can be motivated and argued from different perspectives. The most recent routine construction \cite{Biswas:2016egy,Koshelev:2017tvv} shows that such an action with analytic infinite derivative (AID) form-factors appears as the most general action describing linear fluctuations around maximally symmetric space-times. Analyticity of form-factors is required by the existence of the low-energy GR limit while presence of exactly infinite number of derivatives follows from imposing ghost-free conditions.

The non-locality scale naturally bounded by the Planck mass from above and also is significantly large and close to the Planck mass as it will follow later. As such
$$\Mc_s\lesssim M_p\,.$$
Obviously, the non-locality scale can be associated with more fundamental originating theories like string theory for example.

\section*{From trace anomaly to AID gravity}
Looking back we know that  GR was extremely successful matching predictions of several observations over the century including the very recent detection of gravitational waves \cite{Wald:1984rg,Abbott:2016blz}. 
However, it is well known from the very beginning that GR requires modifications at small distance and time scales or at high energies which are assumed the energies to be comparable with the Planck energy scales. Quantum Field Theory (QFT)  would obviously form an initial guiding path towards understanding rules for modifying gravity \cite{Duff:1993wm}. GR is a classical theory of geometry but it cannot be quantized or made ultraviolet (UV) complete because it is non-renormalizable. In this regard a different question was asked in the 1970's which is {\it can quantum corrections to the geometry arise due to matter fields?} Answer to this question turned out to be one of the most important development known as {\it trace anomaly} which  led to significant developments in classical and quantum gravity \cite{Duff:1993wm}. The trace anomaly represents 1-loop quantum corrections to the graviton propagator due to conformal fields leading to an anomalous non-zero trace of energy momentum tensor which can in short be written in the following form 
\begin{equation}
M_p^2 R = g^{\mu\nu}\langle T_{\mu\nu} \rangle = a F+ bG+c\square R\,,  
\label{tra}
\end{equation} 
where one has
 the Weyl tensor square term $F=W_{\mu\nu\rho\sigma}W^{\mu\nu\rho\sigma}$, the Gauss-Bonnet term $G= R_{\mu\nu\rho\sigma}R^{\mu\nu\rho\sigma}-4 R_{\mu\nu}R^{\mu\nu} +R^2$ with $R_{\mu\nu\rho\sigma}$ being Riemann tensor, $R_{\mu\nu}$ being Ricci tensor and $a,b,c$ are dimensionless coefficients that depend on the number of conformal fields. Trace anomaly has been widely studied and applied in various fundamental theories including anticipated UV complete theories such as string theory and supergravity (SUGRA) \cite{Duff:1993wm}. 
 
 A first and very significant application of trace anomaly appeared in cosmology and led to the discovery of a non-singular solution in the form of Friedmann-Lema{\^i}tre-Robertson-Walker (FLRW) space-time describing an epoch of an exponential (quasi-de Sitter) expansion in the early Universe \cite{Starobinsky:1980te,Hawking:2000bb}, the epoch well known as \textit{the cosmic inflation}. The paradigm of inflation does successfully resolves the standard problems of {\it the big bang cosmology} \cite{Guth:1980zm,Linde:1981mu}. This particular inflationary scenario is the Starobinsky inflation model happening in a regime when the last term in (\ref{tra}) dominates over the rest and this effect can be equivalently seen with the following effective action which we shortly call as (local) $R^2$ gravity 
 \begin{equation}
 S_{R^2}= \int d^4x\sqrt{-g} \LT \frac{M_p^2}{2}R + \frac{M_p^2}{12M^2} R^2 \RT\,. 
 \label{LSR2}
 \end{equation}
where $M\ll M_p$ becomes the mass of a propagating scalar in this model, named scalaron. During inflationary regime the scalaron degree of freedom coincides with the Bardeen potential
$\Psi$ \cite{Bardeen:1980kt} which parametrizes metric fluctuations and whose two point correlation function gives the power spectrum of curvature perturbations $\Rc \approx -\frac{1}{\epsilon} \Psi $  which explains the density or temperature fluctuations observed in the cosmic microwave background (CMB). 
Here $\epsilon = -\frac{\dot{H}}{H^2}\approx\frac{M^2}{6H^2}\ll 1 $ is the so-called slow-roll parameter, $H,\,\dot{H}$ are the Hubble parameter and its  derivative with respect to the cosmic time respectively.


The only free parameter $M$ in this model is fixed to $M\approx 1.3\times 10^{-5}M_p$ by the
observed Fourier power spectrum $\Pc_{\Rc}= 2.2\times 10^{-9}$ of primordial scalar 
perturbations (or matter density perturbations) in the Universe corresponding to $N=55$ number of $e$-foldings during inflation \cite{Akrami:2018odb}. The key predictions of $R^2$ inflation are the spectral index $n_s$
which is the tilt of the power spectrum $P_\Rc$, the ratio of the tensor to the scalar power spectra $r$
and the tensor tilt $n_t$
at the Hubble radius exit. In the local $R^2$ model these parameters become:
\begin{equation}
	n_s = 1-\frac{2}{N}\,,\quad r= \frac{12}{N^2}\,,\quad r=-8n_t\,.
\label{keypr}
\end{equation}
The current observational bounds read \cite{Akrami:2018odb} $ n_s= 0.9649\pm 0.0042\,\, \textrm{at}\,\, 68\% \textrm{CL}, r<0.064 \,\,\textrm{at}\,\, 95\%\, \textrm{CL}$ making $R^2$ inflation incredibly well consistent with the present CMB measurements by PLANCK mission and the more recent BICEP2/Keck Array experiments \cite{Akrami:2018odb}.  The last relation in (\ref{keypr}) implies the negative tensor tilt $n_t<0$ and this has been regarded as a definitive test especially for $R^2$ inflation. 

From the theoretical perspective $R^2$ gravity is proven to be renormalizable by the addition of the Weyl tensor squared term, the idea proposed and studied by Stelle \cite{Stelle:1976gc,Stelle:1977ry} which makes the theory UV complete in the scalar sector but the unitarity gets spoiled by appearance of a spin-2 ghost. Despite there have been several attempts in the last decades of realizing $R^2$ gravity within more fundamental UV complete theories such as string theory and supergravity \cite{Linde:2014nna,Kehagias:2013mya}, there are interesting advances beyond Stelle's theory having certain connections with a manifestly UV complete setup of string field theory \cite{Witten:1985cc,Witten:1986qs,Zwiebach:1985uq,Arefeva:2004odl} and this is our main focus in the sequel. 

Deeper developments of trace anomaly computations resulted in a non-local effective theory of gravity  \cite{Barvinsky:2015,Teixeira:2020kew}. Such a non-local gravity contains quadratic in curvatures terms with infinitely many derivatives in a form of non-analytic functions such as $\ln \frac{\square}{\mu^2}$ \cite{Barvinsky:2015} where $\mu$ is the theory cut-off scale. These non-local gravitational theories could lead to ghost degrees of freedom and moreover do not have a smooth low energy limit towards GR making themselves not interesting as  UV completion candidates of local $R^2$ gravity. Nevertheless, this provides an insight that a non-locality can play a crucial role on the way to UV completion.
Moreover, various approaches towards quantum gravity including causal sets \cite{Belenchia:2014fda}, non-commutative theories \cite{Witten:1985cc,Chu:2005nb}, loop quantum gravity \cite{Hossenfelder:2007re,Ashtekar:2007tv} and asymptotic safety \cite{Weinberg:2009wa,Bonanno:2017pkg,Liu:2018hno,Eichhorn:2018yfc,Bosma:2019aiu} highlight the importance of gravity being non-local near Planck scale.  
	Since non-locality appears to be a common feature in several frameworks aiming to construct quantum gravity, building of a non-local gravity theory with viable inflationary scenario gives us the best opportunity to look for its signatures in the CMB.\footnote{Signatures for non-local nature of gravity has also been studied recently in the context of astrophysical and gravitational wave observations \cite{Capozziello:2017xla,Calmet:2017rxl}. In \cite{Capozziello:2017xla} gravitational energy-momentum pseudo tensor is calculated in the context of non-local theories and its importance with respect to the power carried by gravitational radiation from astrophysical sources has been discussed. In \cite{Calmet:2017rxl}, it was pointed out that one can probe the parameters of the effective action of quantum gravity with non-analytic form factors using a combination of gravitational wave observations and pendulum type experiments searching deviations of Newtonian potential.} 

	In string field theory analytic non-locality does naturally appear because interactions are intrinsically non-local. In this scenario infinite derivative operators like $e^{{\square}/{\Mc_s^2}}$, with $\Mc_{s}$ being a string scale $\Mc_{s}\lesssim M_p$ enter the vertex terms of space-time fields interactions. 
	
	Notably the AID gravity action in (\ref{NC-action}) joins at once several remarkable properties: it manifestly inherits the non-local operator structure from the string theory, this in turn targets questions of the UV completion by the very structure of the action and also this proposal does solve the ghost problem of Stelle gravity \cite{Biswas:2016egy}. For all the aspects presence of AID form-factors is the crux.

\section*{$R^2$-like inflation in AID gravity (\ref{NC-action}) and future CMB}

It was proven that the local $R^2$ gravity  inflationary solution that itself satisfies $\square R= M^2 R$ happens to be exact attractor solution of (\ref{NC-action}) with very minimal conditions such that $\Fc_R\LF \frac{M^2}{\Mc_{s}^2} \RF =\frac{M_p^2}{6M^2}=\Fc_1$ and $\Fc_R^{(1)}\LF \frac{M^2}{\Mc_{s}^2} \RF =0$ (hereafter $^{(1)}$ means the derivative with respect to the argument).  We recall that $M$ is the scalaron mass. No condition arise on $\Fc_W\LF \square_{s} \RF$ at the background level since Weyl tensor is zero in FLRW space time. Study of scalar and tensor degrees of freedom in quasi-de Sitter regime imply the following general structures of form-factors
\begin{equation}
	\Fc_{R}\LF \square_s \RF = \Fc_{1}\frac{3e^{\gamma_S\LF \square_s \RF } \LF \square_s-\frac{M^2}{\Mc_{s}^2} \RF+\LF \frac{\bar{ R}}{\Mc_{s}^2}+3\frac{M^2}{\Mc_{s}^2} \RF}{3\square_s+\frac{\bar{ R}}{\Mc_{s}^2}} \,,\quad \Fc_W\LF \square_s \RF = \frac{\Fc_{1}\bar{ R}}{\Mc_{s}^2}\frac{e^{\gamma_T
		\LF \square_s-\frac{2}{3}\frac{\bar{ R}}{\Mc_{s}^2} \RF}-1}{\square_s-\frac{2\bar{ R}}{3\Mc_{s}^2}}\,.
\label{formfactors2s}
\end{equation}
where $\gamma_S,\,\gamma_T$ are entire functions of the d'Alembertian operator and $\bar R$ is the background curvature of the qusi-de Sitter phase. The structure of form-factors is dictated by the absence of ghosts requirement.
The following heuristic hierarchy
 \begin{equation}
 M^2\ll H_{\text{inf}}^2 \lesssim \Mc_{s}^2 \lesssim M_{p}^2\,, 
 \end{equation}
 where $H_{\text{inf}}^2\approx \bar R/12$ is the Hubble parameter squared during inflation is proven to keep the key well estimated observational predictions of $\Pc_{\Rc},\,n_s$ to remain the same as in local $R^2$ model \cite{Koshelev:2017tvv}. Given the great observational precision in this instances it is crucial to name new finding successful.

\begin{figure}[H]
\centering	\includegraphics[height=1.75in] {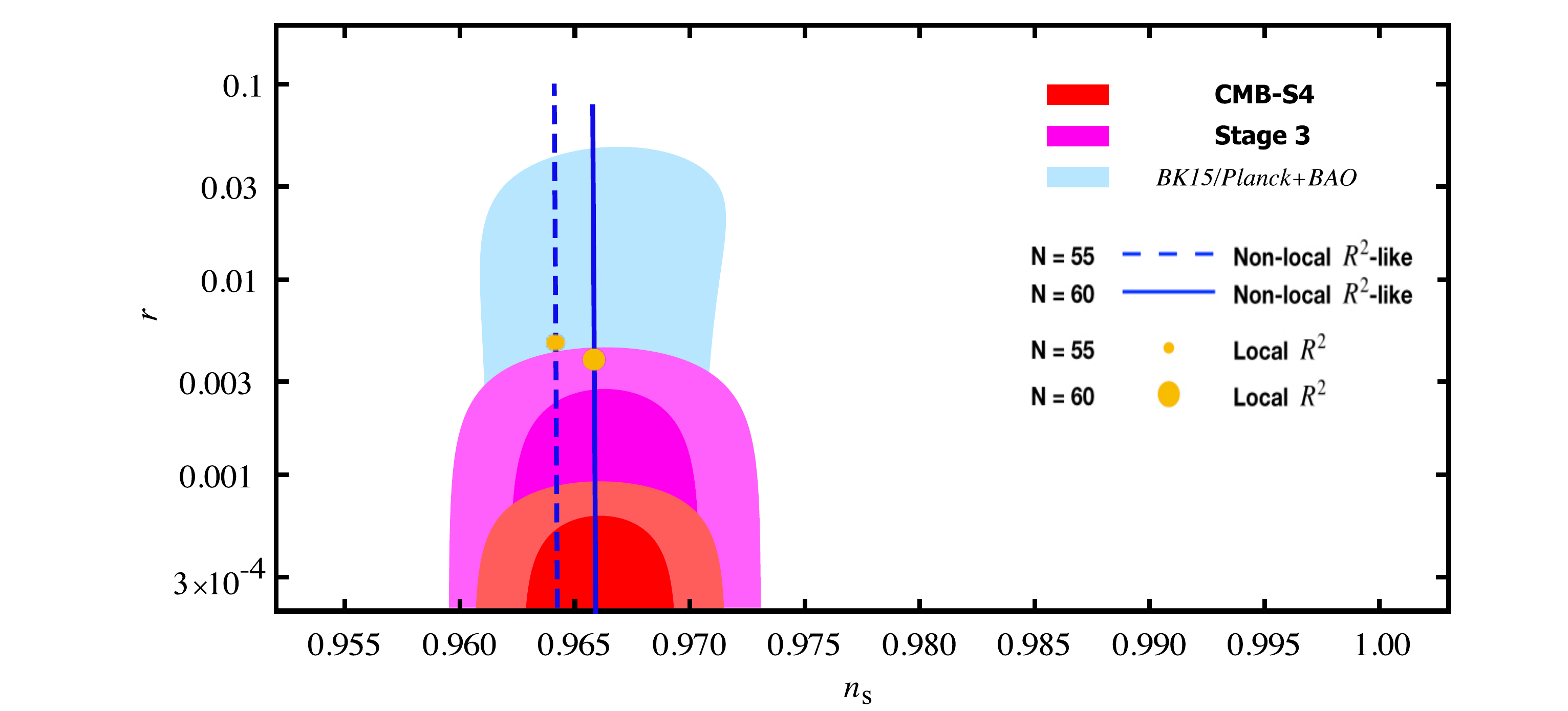},\quad\includegraphics[height=1.75in]{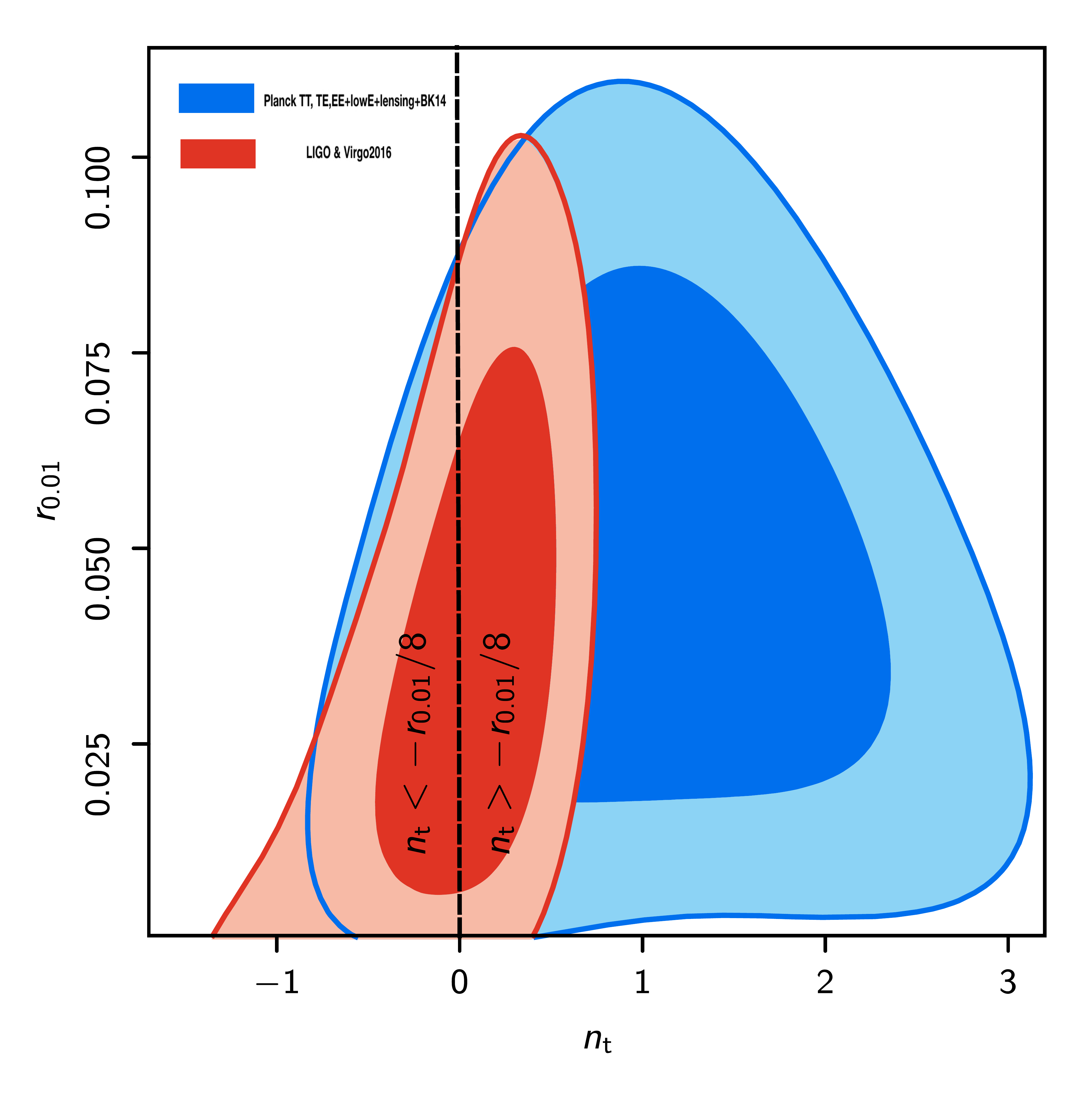}\caption{In the left panel, we depict the predictions of non-local $R^2$-like inflation in the $(n_s,r)$ plane of the latest CMB S-4 science paper about future forecast of detecting $B$-modes \cite{Abazajian:2019eic}. In the right panel, we note that the predictions of non-local $R^2$-like inflation can be anywhere within the likelihood projected $(n_t,r)$ plane of latest Planck 2018  \cite{Abazajian:2016yjj}. }
	\label{Staro-plot-1} 
\end{figure}

 The two new predictions of inflation in AID theory (\ref{NC-action}) \cite{Koshelev:2016xqb,Koshelev:2017tvv,Koshelev:2020foq} which can clearly discriminate local and non-local theory, and many other setups of local $R^2$-like inflationary scenarios \cite{Akrami:2018odb} are 
\begin{equation}
r=\frac{12}{N^{2}}e^{-2\gamma_T\LF\frac{-\bar{R}}{2\Mc_s^{2}}\RF}\Bigg\vert_{k=aH} ,\quad n_{t}\approx-\frac{3}{2N^{2}}-\left(\frac{2}{N}+\frac{3}{2N^{2}}\right)\frac{\bar{R}}{2\mathcal{M}_s^{2}}\gamma_T^{(1)}\left(-\frac{\bar{R}}{2\mathcal{M}_s^{2}}\right)\Bigg\vert_{k=aH}\,,
\end{equation} 
where $\bar{ R}_{k=aH}\approx 4NM^2$ is the value of Ricci scalar  at the Hubble radius crossing.
There are three important things to be noticed here: 
	tensor to scalar ratio can take any value depending on the form-factor of Weyl tensor quadratic term; tensor tilt gets correction due to non-locality in the model and a deviation from the standard consistency relation $r=-8n_t$ is evident here; a possibility of a negative tensor tilt $n_t>0$ is not excluded. 
This means that the non-local $R^2$-like inflation is manifestly ready to be tested by future CMB observations \cite{Abazajian:2019eic} (see Fig.~\ref{Staro-plot-1}).

\section*{Beyond two-point correlations}
\begin{figure}[H]
		\centering\includegraphics[height=1.72in]{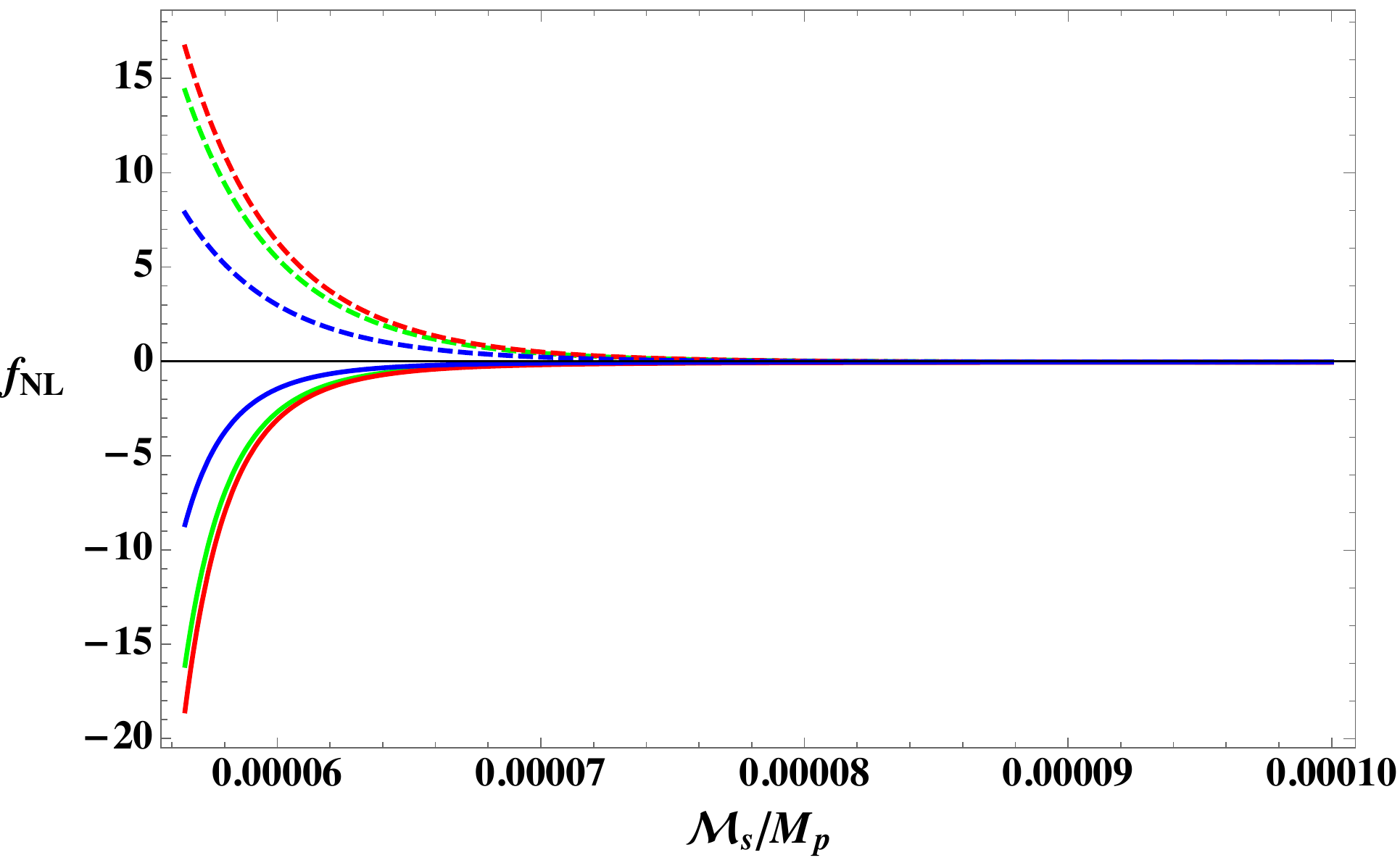}\caption{{In the above plots, $f_{NL}$ versus the scale of non-locality $\Mc_{s}$ (in the units of $M_p$) is depicted for squeezed (blue), equilateral  (red), and orthogonal (green) configurations for the polynomial entire functions $\gamma_S$ given by eqs. (4.19) and (4.20) of \cite{Koshelev:2020foq} and represented by solid and dashed lines respectively. Here $N=55$ of $e$-foldings is assumed. In the limit $\Mc_{s}\to M_p$ the predictions of the local $R^2$ model are recovered.}}
	\label{Staro-plot-fnl} 
\end{figure}

3-point scalar correlations that give non-Gaussian corrections provide another test for the non-local $R^2$-like inflation \cite{Maldacena:2002vr,Meerburg:2019qqi,Meerburg:2016zdz}. 
In the Fourier space, the three-point correlation function of curvature perturbations is defined through
parameter $f_{\text{NL}}$ which is called 
the reduced bi-spectrum. Physically this measures the three-point tree-level interactions of curvature perturbation in quasi-de Sitter regime. There are three triangular configurations which are observationally measurable to constrain the non-Gaussianities in gravity modifications. As per the latest CMB data with respect to Planck, BICEP2/Keck Array and BK15 with TT,TE, EE+lowE+lensing \cite{Akrami:2019izv} we have 
\begin{equation}
	f_{\text{NL}}^{sq}= -0.9\pm 5.1\,,\quad f_{\text{NL}}^{eq}= -26\pm 47\,,\quad f_{\text{NL}}^{ortho}= -38\pm 24\,,~\textrm{at}~68\%~\textrm{CL}. 
\label{lastest.fnl}
\end{equation}
for the so-called squeezed, equilateral, and orthogonal configurations respectively. Among these the squeezed limit which tells us about an interaction between a long wavelength mode that has exited the Hubble radius already with those modes which are still exiting the Hubble radius is what is potentially measurable in near future and any detection of it away from Maldacena's single canonical field consistency relation  would tell us more about inflationary paradigm and UV completion \cite{Maldacena:2002vr,Chen:2006nt,Arkani-Hamed:2015bza}.
This consistency relation is obeyed by local $R^2$ inflationary models and is read
$$f_{\text{NL}}^{sq}=\frac{5}{12}\LF 1-n_s \RF\, .$$
In the case of non-local $R^2$-like inflation the result for the squeezed limit of non-Gaussianities turns out to be \cite{Koshelev:2020foq}
	\begin{equation}
\begin{aligned}
f_{NL}^{sq} \approx  \frac{5}{12} \LF 1-n_s \RF -23 \epsilon^2\left[e^{\gamma_S \left(\frac{\bar{ R}}{4 \Mc_s^2}\right)}-1\right] -\frac{4\bar{ R} }{\Mc_s^2}\epsilon^3 e^{\gamma_S \left(\frac{\bar{ R}}{4 \Mc_s^2}\right)} \gamma _S^{(1)}\left(\frac{\bar{ R}}{4 \Mc_s^2}\right)\Bigg\vert_{k=aH}\,,
\end{aligned}
\label{fnlf}
\end{equation}
which gets correction due to non-local tree level interactions of curvature perturbation. The fact of getting a possibility of $f_{\text{NL}}^{sq} \sim\Oc(1)$ (see Fig.~\ref{Staro-plot-fnl}) due to non-local effects puts this non-local model within the scope of upcoming CMB and astrophysical observations \cite{Meerburg:2019qqi,Meerburg:2016zdz}. Further, the other configurations also get non-local corrections as follow \cite{Koshelev:2020foq}
\begin{equation}
\begin{aligned}
	f_{NL}^{eq} & \approx \frac{5}{12} \LF 1-n_s \RF -49 \epsilon^2\left[e^{\gamma_S \left(\frac{\bar{ R}}{4 \Mc_S^2}\right)}-1\right]-\frac{9 \bar{ R} }{ \Mc_s^2}\epsilon^3 e^{\gamma_S \left(\frac{\bar{ R}}{4 \Mc_s^2}\right)} \gamma _S^{(1)}\left(\frac{\bar{ R}}{4 \Mc_s^2}\right)\Bigg\vert_{k=aH}\,,\\ 
f_{NL}^{ortho} & \approx \frac{5}{12} \LF 1-n_s \RF -43 \epsilon^2\left[e^{\gamma_S \left(\frac{\bar{ R}}{4 \Mc_s^2}\right)}-1\right] -\frac{22\bar{ R} }{3 \Mc_s^2}\epsilon^3 e^{\gamma_S \left(\frac{\bar{ R}}{4 \Mc_s^2}\right)} \gamma _S^{(1)}\left(\frac{\bar{ R}}{4 \Mc_s^2}\right)\Bigg\vert_{k=aH}\,. 
\end{aligned}
\label{fnlf}
\end{equation}

{\it In conclusion, as the immense number of approaches to quantum gravity indicates a non-local nature of gravity, realizing $R^2$-like inflation in a non-local configuration given by AID gravity action (\ref{NC-action}) provides a unique hope to test nature of gravity with CMB observations. Any detection of the predictions of non-local $R^2$-like inflation would imply a giant leap towards our understanding of quantum gravity. }

	\acknowledgments
AK is supported by FCT Portugal investigator project IF/01607/2015. 
This research work was supported by grants  UID/MAT/00212/2019, COST Action CA15117 (CANTATA). KSK  is  supported  by Netherlands Organization for Scientific Research (NWO) grant number 680-91-119.  AAS was partly supported by the project number 0033-2019-0005 of the Russian
Ministry of Science and Higher Education.

\bibliographystyle{utphys}
\bibliography{ssa.bib}

\end{document}